\documentclass[showpacs,aps,graphicx,twocolumn]{revtex4}
\usepackage{amssymb}
\usepackage{graphicx}

\begin{document}

\title{High-capacity quantum secure direct communication based on quantum hyperdense coding with hyperentanglement\footnote{Published in Chin.
Phys. Lett.  \textbf{28}(4), 040305 (2011).}}

\author{Tie-Jun Wang$^{1,2,3}$, Tao Li$^{1}$, Fang-Fang Du$^{1}$, and Fu-Guo Deng$^{1,}$\footnote{Corresponding author:
fgdeng@bnu.edu.cn}}
\address{$^1$ Department of Physics, Applied Optics Beijing Area Major
Laboratory, Beijing Normal University, Beijing
100875\\
$^2$ College of Nuclear Science and Technology, Beijing Normal
University, Beijing 100875\\
$^3$ Beijing Radiation Center, Beijing 100875}

\date{\today }

\begin{abstract}
We present a quantum hyperdense coding protocol with
hyperentanglement in  polarization and  spatial-mode degrees of
freedom of photons first and then give  the details for a quantum
secure direct communication (QSDC) protocol based on this quantum
hyperdense coding protocol. This QSDC protocol has the advantage of
having a higher capacity than the quantum communication protocols
with a qubit system. Compared with the QSDC protocol based on
superdense coding with $d$-dimensional systems, this QSDC protocol
is more feasible as the preparation of a high-dimension quantum
system is more difficult than that of a two-level quantum system at
present.
\end{abstract}

 \maketitle

Quantum mechanics provides some good ways for secure quantum
communication which is originally used to create a private key
between two remote legitimate users, say the sender Alice and the
receiver Bob. The non-cloning theorem provides a physical principle
for the fact that an eavesdropper can not steal the information
transmitted from the sender to the receiver fully and freely as an
unknown quantum state can not be copied perfectly. In 1984, with
four nonorthogonal single-photon polarization states, Bennett and
Brassard \cite{bb84} proposed the first quantum key distribution
(QKD) protocol. In this QKD protocol, Alice prepares her
single-photon states by choosing randomly one of two measuring bases
(MBs), say $Z$ and $X$, and Bob measures his single-photon states by
choosing randomly these two MBs after he receives the photons from
Alice. In this way, Alice and Bob will obtain some correlated
outcomes with the success probability $50\%$. In 1992, Bennett
\cite{b92} simplified this protocol with two nonorthogonal states
and decreased the success probability to 25\% in principle. The
nonlocality feature of an entangled system also provides a good tool
for designing secure quantum communication. For example, in 1991,
Ekert \cite{Ekert91} designed a QKD protocol with
Einstein-Podolsky-Rosen (EPR) pairs, resorting to the Bell
inequality for checking eavesdropping, and its success probability
is also 50\%. Subsequently, Bennett, Brassard, and Mermin
\cite{BBM92} simplified its process  for checking eavesdropping.
Now, QKD has attracted much attention \cite{longqkd,BidQKD,RMP}.

Different from QKD, quantum secure direct communication (QSDC) is
used to transmit the secret message securely and directly, not
resorting to a private key. In 2002, Long and Liu \cite{longqkd}
introduced the concept of quantum data block into quantum
communication and proposed the first QSDC scheme based on entangled
quantum systems and it is detailed in the two-step QSDC protocol
\cite{two-step} in 2003. Of course, in 2002 Beige \emph{et al.}
\cite{BeigeA} proposed another deterministic quantum communication
scheme based on single-photon two-qubit states, in which the message
can be read out only after a transmission of an additional classical
information for each qubit, i.e., the cryptographic key of the
sender \cite{longreview}. In 2002, Bostr\"om and Felbinger
\cite{pingpong} presented a quasi-secure direct communication scheme
with an EPR pair, call it the ping-pong scheme. However, the
ping-pong scheme is insecure as the title suggests, and it can be
attacked fully if the transmission efficiencies are lower than 60\%
as the two parties of quantum communication only exploit one basis
to check eavesdropping \cite{attachpp}. In 2004, Cai \emph{et al.}
\cite{caipra} improved the ping-pong scheme with the similar way to
the two-step protocol
 \cite{two-step} and proposed another deterministic secure
communication protocol based on single photons \cite{cai}. The
quantum one-time pad protocol \cite{QOTP} is the first QSDC scheme
based on a sequence of single photons with a single-qubit quantum
privacy amplification protocol \cite{privacy}. In 2005, Wang
\emph{et al.} \cite{Wangc} proposed a QSDC protocol with
high-dimension superdense coding. Now, there are many deterministic
quantum communication protocols, including QSDC protocols
\cite{longqkd,two-step,QOTP,Wangc,Wangc2,lixhcp} and deterministic
secure quantum communication protocols
\cite{lixhpra,BeigeA,yan,gao2,gao3,
zhangzj,zhangsPRA,lixhjp,Wangc3,wangjpla,gao4,song,gubqsdc}.

Hyperentanglement is a state of being simultaneously entangled in
multiple degrees of freedom and it plays an important role in
quantum information processing recently. For example, it is used to
analyze the four Bell states in polarization degree of freedom of
photons \cite{BSA1,BSA2,BSA3}.  In 2008, with the help of the
hyperentangled state in both polarization and orbit angular
momentum, Barreiro \emph{et al.} \cite{Densecodingimprove} beated
the channel capacity limit for linear photonic superdense coding.
In 2002, Simon and Pan  \cite{Simon} exploited the hyperentanglement
in spatial-mode and  polarization degrees of freedom of photons to
purify the entanglement in polarization degree of freedom. In 2009,
Sheng, Deng and Zhou \cite{shengpraepp1} exploited the
hyperentanglement in spatial and polarization degrees of freedom to
purify a parametric down-conversion  entanglement source. In 2010,
Sheng and Deng \cite{shengpraepp2} proposed the concept of
deterministic entanglement purification and presented the first
protocol with hyperentanglement in polarization, spatial, and
frequency degrees for freedom. In 2010, an one-step  deterministic
polarization entanglement purification protocol by using spatial
entanglement was proposed \cite{shengpraepp3,lixhpraepp} with
hyerpentanglement in polarization and spatial-mode degrees of
freedom. More recently, a complete hyperentangled-Bell-state
analysis (CHBSA) protocol for quantum communication was proposed
with the help of cross-Kerr nonlinearity \cite{shengBell}. With
CHBSA, some important processes can be accomplished such as
teleportation and entanglement swapping in two degrees of freedom,
which will improve the capacity of quantum communication largely.

In this Letter, we will first introduce  a quantum hyperdense coding
protocol with hyperentanglement in  polarization and  spatial-mode
degrees of freedom of photons, and give the way for encoding the
information with linear optical elements. Then we will give a QSDC
protocol based on this quantum hyperdense coding protocol and
discuss its security against Trojan horse attack strategies.
Compared with the QSDC protocols with a qubit system
\cite{two-step,Wangc2}, this QSDC protocol has the advantage of
having a higher capacity as each photon can carry 4 bits of
information from the sender to the receiver. Compared with the QSDC
protocol based on superdense coding with $d$-dimension systems
\cite{Wangc}, this QSDC protocol is more feasible as the preparation
of a high-dimension quantum system is more difficult than that of a
two-level quantum system at present.

\begin{figure}[!ht]
\begin{center}
\includegraphics[width=6cm,angle=0]{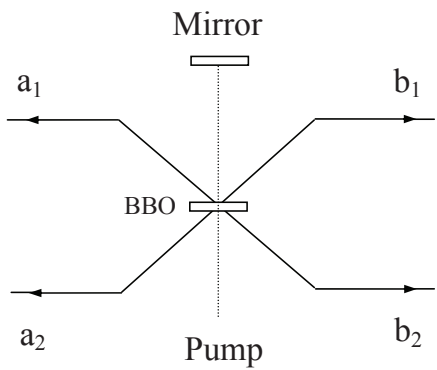}
\caption{Schematic diagram of a quantum hyperentanglement source in
 polarization and  spatial-mode degrees of freedom, which can be
produced with a beta barium borate (BBO) crystal and a pump pulse of
ultraviolet light.}\label{fig1_entanglement}
\end{center}
\end{figure}

A hyperentangled state in  polarization and  spatial-mode degrees of
freedom can be described as follows:
\begin{eqnarray}
|\Phi^{+}_{AB}\rangle_{PS}=\frac{1}{2}(|HH\rangle+|VV\rangle)_{AB}\otimes(|a_1b_1\rangle
+ |a_2b_2\rangle)_{AB},\label{hyperentanglement}
\end{eqnarray}
where $H$ and $V$ are the horizontal and the vertical polarizations
of photons, respectively. The subscripts $A$ and $B$ represent the
two photons in the hyperentangled state. $a_1(b_1)$ and $a_2(b_2)$
are the different spatial modes for photon $A$ ($B$), shown in
Fig.1. The subscript $P$ denotes the polarization degree of freedom
and $S$ is the spatial-mode degree of freedom. As pointed out in
Refs.\cite{Simon,shengpraepp1}, a pump pulse of ultraviolet light
passes through a beta barium borate (BBO) crystal and produces
correlated pairs of photons into the modes $a_{1}$ and $b_{1}$. Then
it is reflected and traverses the crystal a second time, and
produces correlated pairs of photons into the modes $a_{2}$ and
$b_{2}$. The Hamiltonian can be approximately described as
\begin{eqnarray}
H_{PDC} &=& \gamma[(a^{+}_{1H}b^{+}_{1H}+a^{+}_{1V}b^{+}_{1V}) \nonumber\\
&+& r e^{i\phi}(a^{+}_{2H}b^{+}_{2H}+a^{+}_{2V}b^{+}_{2V})]
     + H.c,
\end{eqnarray}
where  $r$ denotes the relative probability of emission of photons
into the lower modes compared to the upper modes, and $\phi$ is the
phase between these two possibilities \cite{Simon,shengpraepp1}. As
the same as Refs. \cite{Simon,shengpraepp1}, in a simple case we
assume $r=1$ and $\phi=0$. Thus the  state of the photon pair can be
described by
$(a^{+}_{1H}b^{+}_{1H}+a^{+}_{1V}b^{+}_{1V}+a^{+}_{2H}b^{+}_{2H}+a^{+}_{2V}b^{+}_{2V})|0\rangle$.
It  also can be written as
$(|a_{1}\rangle|b_{1}\rangle+|a_{2}\rangle|b_{2}\rangle)(|H_{a}\rangle|H_{b}\rangle
+ |V_{a}\rangle|V_{b}\rangle)$. If the pump pulse is faint, the
four-photon state produced by this PDC source  is negligible.

For a quantum system with two photons  hyperentangled in
polarization and  spatial-mode degrees of freedom, there are 16
generalized Bell states, shown as
\begin{eqnarray}
|\Psi_{AB}\rangle_{PS}=\vert \theta \rangle_P \otimes \vert \xi
\rangle_S, \label{hyperentanglement}
\end{eqnarray}
where $\vert \theta \rangle_P$ is one of the four Bell states  in
polarization degree of freedom as
\begin{eqnarray}
|\phi^{\pm}\rangle_{P}=\frac{1}{\sqrt{2}}(|HH\rangle\pm|VV\rangle)_{AB},\nonumber\\
|\psi^{\pm}\rangle_{P}=\frac{1}{\sqrt{2}}(|HV\rangle\pm|VH\rangle)_{AB},\label{polarization}
\end{eqnarray}
and $\vert \xi \rangle_S$ is one of the four Bell states in
spatial-mode degree of freedom as
\begin{eqnarray}
|\phi^{\pm}\rangle_{S}=\frac{1}{\sqrt{2}}(|a_1b_1\rangle \pm |a_2b_2\rangle)_{AB},\nonumber\\
|\psi^{\pm}\rangle_{S}=\frac{1}{\sqrt{2}}(|a_1b_2\rangle \pm
|a_2b_1\rangle)_{AB}.\label{spatial}
\end{eqnarray}
With CHBSA \cite{shengBell}, one can in principle distinguish these
16 hyperentangled Bell states $|\Psi_{AB}\rangle_{PS}$.

Now, we first construct a quantum hyperdense coding protocol with
the hyperentanglement in  polarization and  spatial-mode degrees of
freedom and then apply it into QSDC for obtaining a high channel
capacity.

Similar to the quantum dense coding protocols
\cite{densecoding,superdensecoding,superdensecoding2} with
entanglement in  polarization degree of freedom, a quantum
hyperdense coding protocol can be accomplished with the following
steps:

(1) The receiver Bob prepares a hyerentangled photon pair $AB$ in
the state $|\Phi^{+}_{AB}\rangle_{PS}$, as shown in
Fig.\ref{fig1_entanglement}.

(2) Bob sends the photon $A$ to the sender Alice with two fiber
channels.

(3) After receiving the photon $A$, Alice encodes her information on
the photon $A$ with the unitary operation $U_{ij}=U^P_i\otimes
U^S_j$ ($i,j=1,2,3,4$). Here
\begin{eqnarray}
U^P_1 &=&   \vert H\rangle\langle H\vert + \vert V\rangle\langle
V\vert, \;\;\;\;\;\; U^P_2 =   \vert H\rangle\langle H\vert - \vert
V\rangle\langle V\vert, \nonumber\\
U^P_3 &=&    \vert V\rangle\langle H\vert + \vert H\rangle\langle
V\vert, \;\;\;\;\;\; U^P_4  = \vert V\rangle\langle H\vert - \vert
H\rangle\langle V\vert, \nonumber\\
U^S_1 &=& \vert a_1\rangle\langle a_1\vert + \vert a_2\rangle\langle
a_2\vert, \;\; U^S_2 =  \vert a_1\rangle\langle a_1\vert - \vert
a_2\rangle\langle a_2\vert, \nonumber\\
U^S_3 &=&  \vert a_2\rangle\langle a_1\vert + \vert
a_1\rangle\langle a_2\vert, \;\; U^S_4 = \vert a_2\rangle\langle
a_1\vert - \vert a_1\rangle\langle a_2\vert.
\end{eqnarray}

(4) After performing one of the 16 unitary operations on the photon
$A$, Alice sends it to Bob.

(5) Bob distinguishes the state of the hyperentangled photon pair
$AB$ with CHBSA \cite{shengBell} and can obtain 4 bits of
information from Alice.

With these five steps, Alice and Bob can in principle accomplish
their quantum hyperdense coding. In fact, this quantum hyperdense
coding protocol is just the generation of the original quantum dense
coding protocol from a two-level systems to a hyperentangled system
with two degrees of freedom. Also, it can be viewed as an
experimental example for the quantum superdense coding protocol with
a $d$-level quantum system \cite{superdensecoding,Wangc}.  From the
view of quantum entanglement source, this hyperdense coding protocol
may be a feasible one with current techniques as the
hyperentanglement in  polarization and spatial degrees of freedom of
photons can be prepared with a BBO crystal at experiment. The four
unitary operations in the polarization degree of freedom of photons
can be implemented with linear optical wave plates. The bit-flip
operation $U^S_3$ and  the phase-flip operation $U^S_2$ in the
spatial-mode degree of freedom of photons can be accomplished by
exchanging the two fiber channels ($a_1$ and $a_2$) and by adding a
$\lambda/2$ wave plate on the spatial mode $a_2$, respectively.

\begin{figure}[!ht]
\begin{center}
\includegraphics[width=6cm,angle=0]{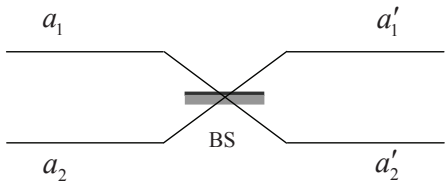}
\caption{Schematic diagram of a Hadamard operation in spatial-mode
degree of freedom with a 50:50 beam splitter (BS).}\label{fig2}
\end{center}
\end{figure}

In quantum communication protocols, Alice and Bob should exploit at
least two uncommutative MBs to check eavesdropping \cite{lixhpra}.
The two MBs in  polarization degree of freedom can be chosen as
$Z^P=\{\vert H\rangle, \vert V\rangle\}$ and $X^P=\{\vert +
\rangle_P=\frac{1}{\sqrt{2}}(\vert H\rangle + \vert V\rangle), \vert
- \rangle_P=\frac{1}{\sqrt{2}}(\vert H\rangle - \vert V\rangle)\}$.
These two MBs can be transferred into each other with a $\lambda/4$
wave plate which acts as a Hadamard opertion. The two MBs in
spatial-mode degree of freedom can be chosen as $Z^S=\{\vert
a_1\rangle, \vert a_2\rangle\}$ and $X^S=\{\vert
+\rangle_S=\frac{1}{\sqrt{2}}(\vert a_1\rangle + \vert a_2\rangle),
\vert -\rangle_S=\frac{1}{\sqrt{2}}(\vert a_1\rangle - \vert
a_2\rangle)\}$. The Hadamard operation between these two MBs can be
completed with a 50:50 beam splitter, shown in Fig.\ref{fig2}. In
order to distinguish the four orthogonal states $\{\vert
H\rangle\vert a_1\rangle, \vert V\rangle\vert a_1\rangle, \vert
H\rangle\vert a_2\rangle, \vert V\rangle\vert a_2\rangle\}$ in
polarization and spatial-mode degrees of freedom of each photon,
Alice should add a polarization beam splitter (PBS) at each of the
two pathes $a_1$ and $a_2$.

With this quantum hyperdense coding protocol, the principle of our
high-capacity  QSDC protocol with hyperentanglement can be described
as follows.

(1) The sender Alice and the receiver Bob agree that the 16 unitary
operations $U_{ij}=U^P_i\otimes U^S_j$ are encoded into 16 different
binary strings.

(2) Bob prepares a sequence of  photon pairs in the hyperentangled
state $|\Phi^{+}_{AB}\rangle_{PS}$. He takes the photon $A$ from
each hyperentangled photon pair $AB$ for making up of an ordered
partner particle  sequence $S_A$, and remaining partner particles
compose another particle sequence $S_B$,  similar to Refs.
\cite{longqkd,two-step,Wangc}.

(3) Bob sends the sequence $S_A$ to  Alice and he keeps the sequence
$S_B$.

(4) After receiving the sequence $S_A$, Alice and Bob check
eavesdropping for the security of their transmission.

The procedure for eavesdropping check can be similar to that in the
two-step QSDC scheme \cite{two-step}: (a) Alice chooses randomly
some sample photons from the  sequence $S_A$ and uses one of two MBs
in each degree of freedom to measure the sample photons randomly.
(b) Alice tells Bob the positions of the sample photons and the
information of the measurements including the outcomes and the MBs.
(c) Bob takes a suitable measurement on each sample photon with the
same MBs as those of Alice's. (d) Bob compares his outcomes with
Alice's to determine whether there is an eavesdropper, say Eve
monitoring the transmission over the two fiber channels. It is the
first eavesdropping check. If their outcomes are correlated, they
can continue the QSDC to next step, otherwise they abort the quantum
communication.

(5) If Alice and Bob confirm that their transmission is secure,
Alice encodes the secret message on the sequence $S_A$ with the
unitary operations $U_{ij}=U^P_i \otimes U^S_j$ and then transmits
the sequence $S_A$ to Bob.

Similar to Refs. \cite{two-step,Wangc}, Alice has to add a small
trick in the procedure of encoding the secret message for the second
eavesdropping check. She will select randomly some photons in the
sequence $S_A$, called sample pairs (which are composed of the
sample photons in the sequence $S_A$ and the photons correlated in
the sequence $S_B$.) and perform on them one of the 16 unitary
operations $U_{ij}$ randomly. It equals to the fact that Alice adds
some redundancy in the coding for checking the security of the
transmission of the  sequence $S_A$ from Alice to Bob. She keeps the
secret including the positions of the sample photons and the
operations performed on them until Bob receives the sequence $S_A$.

(6)  Bob can read out the message encoded on each hyperentangled
photon pair with CHBSA \cite{shengBell}.

(7) Alice tells Bob the positions of the sample pairs and the
unitary operations on them. Bob completes the second
eavesdropping-check analysis.

(8) If the transmission is secure, Bob will obtain the secret
message encoded by Alice. Otherwise, Alice and Bob abandon the
results of the transmission and repeat the procedures from the
beginning.

If this protocol is secure for Trojan horse attack strategies, Alice
and Bob can exploit the quantum correlation between the two photons
in a hyperentangled state to analysis the error rate of the qubits
transmitted and determine whether this protocol is secure for
quantum direct communication or not, similar to the two-step QSDC
protocol \cite{two-step} whose security is in principle the same as
the BBM92 QKD protocol \cite{BBM92,proof,proof2}.

As discussed in Ref. \cite{lixhpra}, in order to prevent Eve from
eavesdropping the quantum communication with Trojan horse attack
strategies, Alice should not only estimate the number of photons in
each legitimate quantum signal in the sequence $S_A$ transmitted
from Bob but also filter out the illegitimate photon signal with
some special filters. The primary Trojan horse attack strategies
include a multi-photon-signal attack \cite{dengattack}, an
invisible-photon attack \cite{caiattack}, or a delay-photon attack
\cite{lixhpra}. At least, these three kinds of Trojan horse attack
strategies in principle do not increase the error rate in the
outcomes obtained by Alice and Bob, and can not be detected with
common eavesdropping check strategies. Certainly, Alice can also
exploit the ways discussed in Ref.\cite{lixhpra} to prevent Eve from
eavesdropping the transmission with these three kinds of Trojan
horse attack strategies. In detail, after receiving the sequence
$S_A$, Alice first adds a filter before the devices with which she
operates the photons including performing unitary operations and
measurements. The filters can only let the wavelengths close to the
operating one in the devices. In this way, the invisible photons in
each quantum signal will be filtered out. Second, Alice exploits a
photon number splitter (PNS) in each path to divide each signal into
two pieces. With the PNS and two single-photon measurements for each
original signal, the users can distinguish whether there exists a
multi-photon signal (including the delay-photon signal and the
invisible photon whose wavelength is so close to the legitimate one
that it cannot be filtered out with the filter) \cite{lixhpra} or
not. At present,  an ideal PNS is not feasible and the users can use
a photon beam splitter (50/50) to replace a PNS for preventing Eve
from stealing the secret message with multi-photon signal attacks.
That is to say, this QSDC protocol based on hyperdense coding is in
principle secure.

Compared with the two-step QSDC protocol \cite{two-step}, each
hyperentangled photon pair in this QSDC protocol can carry 4 bits of
information with one photon traveling forth and back from Bob to
Alice, which means the capacity of this protocol is double of that
in the two-step protocol. Moreover, this QSDC protocol only exploits
an entangled two-level photon system to carry out a high-capacity
quantum communication, not a 4-level photon system. It will make
this QSDC protocol more feasible than the QSDC protocol based
superdense coding \cite{Wangc} as the preparation of a
high-dimension quantum signal is difficult at present
\cite{preparsion}.

In summary, we have introduced a quantum hyperdense coding protocol
with hyperentanglement in  polarization and  spatial-mode degrees of
freedom of photons and then given the details for a quantum secure
direct communication protocol based on this quantum hyperdense
coding protocol. This QSDC protocol has the advantage of having a
higher capacity than the quantum communication protocols with a
qubit system \cite{two-step,Wangc2} as each photon can carry 4 bits
of information from the sender to the receiver. Compared with the
QSDC protocol based on superdense coding with $d$-dimension systems
\cite{Wangc}, this QSDC protocol is more feasible as the preparation
of a high-dimension quantum system is more difficult than that of a
two-level system at present. We also discuss this QSDC protocol
under Trojan horse attack strategies.

This work is supported by the National Natural Science Foundation of
China under Grant No 10974020, the Beijing Natural Science
Foundation under Grant No 1082008, and the Fundamental Research
Funds for the Central Universities.

\end{document}